\documentclass{article}
\usepackage[english]{babel}

\usepackage[utf8]{inputenc}

\usepackage{amsmath, amssymb}
\usepackage{graphicx}
\usepackage{pgfplots}
\usepackage{listings}
\usetikzlibrary{calc}
\usepackage{todonotes}
\usepackage{authblk}
\usepackage{url}
\usepackage{booktabs}
\PassOptionsToPackage{table}{xcolor}
\usepackage{xcolor}
\usepackage{colortbl}     % optional, for \rowcolor, \columncolor, etc.

\usepackage{siunitx}
\usepackage{graphicx}

\newcommand\sconcat{\operatorname{concat}}
\newcommand\sprefixof{\operatorname{prefixof}}
\newcommand\scontains{\operatorname{contains}}
\newcommand\ssuffixof{\operatorname{suffixof}}
\newcommand\sindexof{\operatorname{indexof}}
\newcommand\sat{\operatorname{at}}
\newcommand\ssubstr{\operatorname{substr}}
\newcommand\sstrtoint{\operatorname{str\_to\_int}}
\newcommand\sinttostr{\operatorname{int\_to\_str}}
\newcommand\sreplace{\operatorname{replace}}
\newcommand\sreplaceall{\operatorname{replaceall}}
\newcommand\sreverse{\operatorname{reverse}}
\newcommand\slength{\operatorname{length}}

\newcommand\OMIT[1]{}

\lstdefinelanguage{JavaScript}{
	keywords={typeof, new, true, false, catch, function, return, null, catch, switch, var, if, in, while, do, else, case, break},
	keywordstyle=\color{blue}\bfseries,
	ndkeywords={class, export, boolean, throw, implements, import, this},
	ndkeywordstyle=\color{darkgray}\bfseries,
	identifierstyle=\color{black},
	sensitive=false,
	comment=[l]{//},
	morecomment=[s]{/*}{*/},
	commentstyle=\color{purple}\ttfamily,
	stringstyle=\color{red}\ttfamily,
	morestring=[b]',
	morestring=[b]"
}

\lstdefinelanguage{SMTLib}{
	morekeywords={
		set-logic,declare-fun,assert,check-sat,get-model,
		str.++,str.in.re,str.to.re,re.+,re.union
	},
	sensitive=true,
	morecomment=[l]{;},
	morestring=[b]"
}
\usetikzlibrary{shapes,automata,positioning,decorations.markings, arrows.meta, fit,backgrounds, matrix, fit}
\tikzstyle{block} = [rectangle, draw, text centered, minimum height=1cm, minimum width=3cm]
\tikzstyle{arrow} = [thick,->,>=stealth]
\tikzstyle{bigbox} = [draw, thick, rounded corners, inner sep=0.5cm]
\usepackage{lipsum} % For placeholder text
\usepackage[ND,SEQ,EQ,ML]{prftree}
\title{OSTRICH2: Solver for Complex String Constraints}

\author[1]{Matthew Hague}
\author[2]{Denghang Hu}
\author[3]{Artur Je\.z}
\author[4]{Anthony W. Lin}
\author[5]{Oliver Markgraf}
\author[6]{Philipp R\"ummer}
\author[2]{Zhilin Wu}

\affil[1]{Department of Computer Science, Royal Holloway, University of London}
\affil[2]{Key Laboratory of System Software, Institute of Software, Chinese Academy of Sciences}
\affil[3]{Institute of Computer Science, University of Wroc\l{}aw}
\affil[4]{University of Kaiserslautern-Landau and Max-Planck Institute for Software Systems}
\affil[5]{University of Kaiserslautern-Landau}
\affil[6]{University of Regensburg and Uppsala University}
\date{} 
\begin{document}

	\maketitle

	\begin{abstract}
			We present OSTRICH2, the latest evolution of the SMT solver
            OSTRICH for string constraints. OSTRICH2
            supports a wide range of complex functions on strings and provides
            completeness guarantees for a substantial fragment of
            string constraints, including
            the straight-line fragment and the chain-free fragment.
            OSTRICH2 provides full support for the SMT-LIB theory of
            Unicode strings, extending the standard with several
            unique features not found in other solvers: among others,
            parsing of ECMAScript regular expressions (including look‐around
            assertions and capture groups) and handling of
            user‐defined string transducers.
            We empirically demonstrate that OSTRICH2 is competitive to other 
            string solvers on SMT-COMP benchmarks.
        \OMIT{
            We show by case studies and experiments that OSTRICH2 is not only capable of 
            solving string constraints involving complex
            regular expressions and transducers that lie beyond
            the reach of existing solvers, but also remains competitive on
            standard benchmarks.
        }
            %The solver won SMT-Competition Track QF\_S in
            %year 2023.
            %, which won SMT-Competition'23 Track QF\_S
	\end{abstract}

	\section{Introduction}\label{sec:introduction}
	% !TEX root = main.tex

Strings are ubiquitous in modern software systems, especially with the
advent of programming languages like JavaScript, PHP, and Python. Despite this, 
string manipulation is well-known to be error-prone and can easily lead to 
security vulnerabilities including HTML injection (e.g.\ see \cite{Kern14,LB16}).
Applications to analysis of security vulnerabilities caused by string
manipulation have been one of the main catalysts for the extensive research into
SMT over strings spanning across the last twenty years
\cite{LB16,DBLP:conf/cav/AbdullaACHRRS15,ostrich,Berkeley-JavaScript,HAMPI,BTV09,string-survey}.
One of the success stories of string solvers includes their usage at AWS
for analysis of Role-Based Access Control (RBAC) policies (e.g.\ see
\cite{neha,Cook18}).
Since 2020 SMT-LIB Unicode String theory \cite{SMTLIB} has been formalized and supported by
many existing string solvers, including Z3 \cite{z3}, Z3-alpha \cite{z3alpha}, Z3-Noodler \cite{noodler,noodler-int,noodler-len, noodler-tool}, Z3str3RE \cite{z3str3re},
cvc5 \cite{cvc5}, Z3str4 \cite{z3str4}, Trau \cite{trau0,trau-tool,trau2}, and our own solver OSTRICH \cite{ostrich}.

In this paper, we present OSTRICH2, the latest evolution of the OSTRICH string solver~\cite{ostrich}. The primary contribution of this paper is a complete and unified description of OSTRICH2’s architecture, algorithms, and implementation as a modular and extensible system. Over the years, OSTRICH has undergone extensive internal development such as integrating new solving techniques, heuristics, and architectural changes, but many of these advances have not been previously documented or published. This includes, for instance, the use of word equation splitting strategies, character-count and length abstractions in preprocessing, and a portfolio-based orchestration of multiple solving engines. OSTRICH2 also includes a newly introduced solver based on algebraic data types (ADT-Str), described here for the first time. 
This paper is the first to present the overall system design of OSTRICH2, detailing how its components interact within a modular proof framework.
We also provide the first full account of OSTRICH2’s preprocessing strategies and rule infrastructure. 
These advances have radically enhanced the solver and resulted in significant performance improvements; notably, OSTRICH achieved first place in the QF\_S track of the SMT-COMP 2023 competition.
Together, this constitutes the first comprehensive system description of OSTRICH2.
To this end, rather than being exhaustive, we aim for 
accessibility by taking the reader through illustrative examples (among others).
In addition to unravelling the design of OSTRICH2, we also report our latest
experimentation with the solver on SMT-LIB'25 benchmarks, showing
its competitiveness, most notably on unsatisfiable instances.

The following is a list of main features  of OSTRICH2:
\begin{description}
    \item[(F1)] SMT-LIB Unicode String theory inputs.
\item[(F2)] Native support of features of ECMAScript regular expressions 
(e.g., lookaround assertions, capture groups, and references; 
back-references are supported only in the replacement string of 
\texttt{replace} and \texttt{replace\_all}, not in the matching expression).
    \item[(F3)] Native support of \texttt{replace\_all} and, more generally, complex string
        transductions using a more comprehensive \emph{regular constraint
        propagation} strategy, which \emph{combines} backward and
        forward regular propagations for a plethora of complex functions.
    \item[(F4)] An extensive portfolio of solving strategies that includes
        a cost-enriched string solver (CE-Str) \cite{atva2020}, a list-based ADT 
        solver (ADT-Str), 
        and preprocessings (length/character-count abstraction).
\end{description}

\paragraph*{Related Tools.}
Over the past decade, numerous string solvers have been proposed, employing techniques such as bit-vector encodings, automata-based propagation, and reductions to word equations. Several early or now-unsupported tools do not handle SMT-LIB 2.6 and are no longer actively maintained, including HAMPI~\cite{HAMPI}, Kaluza~\cite{kaluza}, STRANGER~\cite{stranger}, S3~\cite{S3}, Norn~\cite{DBLP:conf/cav/AbdullaACHRRS15}, Trau~\cite{trau2}, and members of the Z3Str family such as Z3str3~\cite{z3str3} and Z3str3RE~\cite{z3str3re}. These solvers contributed important techniques such as word-equation reasoning with advanced heuristics and direct regular expression support, but have not appeared in recent SMT-COMP competitions and are not included in our evaluation.

Our comparisons focus on state-of-the-art, actively maintained SMT solvers with SMT-LIB 2.6 support. Z3-alpha~\cite{z3alpha}, which builds on Z3str4~\cite{z3str4} and Z3~\cite{z3}, remains actively maintained and incorporates strategy synthesis over multiple string-solving backends. Z3~\cite{z3} and cvc5~\cite{cvc5} combine rewriting, word-equation decomposition, and regular constraint reasoning. 
Z3-Noodler~\cite{noodler} introduces a stabilization algorithm that propagated information on regular constraints to a selected word equation until the inferred regular languages on both sides stabilize. The solver is complete for the chain-free fragment but does not support string transductions or \texttt{replace\_all}.

The original OSTRICH solver~\cite{ostrich} implemented a backward regular constraint propagation strategy, providing completeness for the straight-line fragment and support for transductions. Subsequent extensions added the cost-enriched solver CE-Str~\cite{atva2020} and ECMAScript-style regex support~\cite{popl22}. OSTRICH2, described in this paper, is the first to combine these capabilities with additional preprocessing, a newly implemented ADT-based solver, and a portfolio orchestration of multiple engines, as detailed in Section~\ref{sec:architecture}.

\paragraph*{Organization} Section~\ref{sec:grammar} reviews the SMT-LIB 
Unicode Strings standard and our grammar; Section~\ref{sec:architecture} reviews the architecture of OSTRICH2; Section~\ref{sec:string-theory-algorithms} details OSTRICH2’s key algorithms; Section~\ref{sec:completeness} discusses the completeness of the algorithms; Section~\ref{sec:extensibility} shows that OSTRICH2 is extensible with user-defined string functions; Section~\ref{sec:experiments} reports evaluation results; and Section~\ref{sec:conclusion} concludes with directions for future work.

	\section{Specification Language}\label{sec:grammar}
	% !TeX root = main.tex

\subsection{SMT-LIB Standard for Unicode Theory of Strings}%

We begin with the input language of OSTRICH2\@.
It is based on SMT-LIB 2.6~\cite{SMTLIB} --- in particular, including support 
for the SMT-LIB Unicode theory and Linear Integer Arithmetic constraints
--- but additionally also supports some ``complex'' string functions including
transductions.
We start by illustrating the supported language features by example and defer the formal description of the grammar to the end of the section.

\subsubsection{SMT-LIB Constraints}

Below is a minimal SMT‑LIB script to show some core features.
The script uses the quantifier-free string theory with linear integer arithmetic (\verb+QF_SLIA+), contains three string variables \verb+x+, \verb+y+, and \verb+z+, and one integer variable \verb+l+.
It \verb+assert+s constraints on the variables that are explained below.
The final statements check the satisfiability of the constraints and produce a satisfying assignment.

\begin{lstlisting}[language=SMTLib,frame=single]
(set-option :produce-models	true)	
(set-logic QF_SLIA)

(declare-fun x () String)
(declare-fun y () String)
(declare-fun z () String)
(declare-fun k () Int)

(assert (= (str.len x) (+ k 1)))
(assert (= x (str.++ y z)))
(assert (str.in_re x (re.+ (re.union (str.to_re "a") (str.to_re "b")))))

(check-sat)
(get-model)
\end{lstlisting}

Constraints are written in a Lisp-style prefix notation.
The first \verb+assert+ requires that the length of \verb+x+ is equal to $\mathtt{k} + 1$.
The second requires that \verb+x+ is equal to the concatenation of \verb+y+ and \verb+z+.
The third requires that \verb+x+ belongs to the regular language
$(\{\mathtt{a}\} \cup \{\mathtt{b}\})^+$,
which represents non-empty sequences of $\mathtt{a}$ and $\mathtt{b}$ characters.

Standard Boolean connectives are supported.
The example below requires \verb+x+ to either be the concatenation of \verb+y+ and \verb+z+ or the concatenation of \verb+z+ and \verb+y+.
\begin{lstlisting}[language=SMTLib,frame=single]
(assert (or (= x (str.++ y z))) (= x (str.++ z y))))
\end{lstlisting}

\subsection{Extensions Beyond the SMT-LIB Standard}

OSTRICH2 provides additional features to model string constraints in practical applications.
A simple example is the \verb+str.reverse+ function.
Below, we introduce the transducer, regular expression, and automata extensions.

\subsubsection{Transducer-based Operations}

OSTRICH2 supports transducers and prioritised finite-state transducers~\cite{BDM14,BM17,popl22}.

A transducer (a.k.a. a \emph{rational transducer}) takes a single string as 
input and produces a single output string.
They enable operations such as HTML encoding (e.g.\ replacing \verb+&+ with \verb+&amp;+) to be defined.
Transducers are written as recursive functions with two arguments: the input and the output.
The example below shows a \verb+toUpper+ transduction.
The base case (line~4) applies when both strings are empty.
The recursive case (lines~5--11) asserts three conditions.
First, both arguments are non-empty.
Second, the head of the output (\verb+y+) is equal to the upper case version of the head of the input (\verb+x+).
Third, the tails of the two arguments recursively satisfy \verb+toUpper+.
To obtain the upper case version of a character, arithmetic is
performed on the character codes. In general, a transducer definition
can
contain multiple mutually recursive functions. The option
\verb!:parse-transducers! instructs OSTRICH2 to translate the recursive
function definition to an internal transducer representation and apply
an automata-based decision procedure to solve the constraints in lines~16--17.

\begin{lstlisting}[language=SMTLib,frame=single]
(set-option :parse-transducers true)

(define-fun-rec toUpper ((x String) (y String)) Bool
(or (and (= x "") (= y ""))
  (and (not (= x "")) (not (= y ""))
       (= (char.code (str.head y))
          (ite (and (<= 97 (char.code (str.head x)))
                    (<= (char.code (str.head x)) 122))
               (- (char.code (str.head x)) 32)
               (char.code (str.head x))))
       (toUpper (str.tail x) (str.tail y)))))

(declare-fun x () String)
(declare-fun y () String)

(assert (= x "Hello World"))
(assert (toUpper x y))
\end{lstlisting}

Prioritised transducers additionally allow certain transitions to take
precedence over other transitions.
That is, a non-deterministic transduction may only succeed on one path if another cannot succeed.
Such transducers are used internally in regular expression handling,
described below, but can also be formulated using the
\verb!define-fun-rec! notation (not shown here). We discuss these below after we
cover extended regular expressions.

\subsubsection{Extended Regular Expression Support}

The SMT-LIB standard defines a minimal set of regular expression
constructs,
all of which are supported by OSTRICH2.
In addition, OSTRICH2 can also parse ECMAScript regular expressions,
following the 2020 version~\cite{ECMAScript11}, and supports regular
expression features such as look-arounds that cannot be directly
expressed in SMT-LIB.
A simple usage of ECMAScript regular expressions is shown below, using the \verb+re.from_ecma2020+ function.
The expression matches either sequences of characters or sequences of digits.
\begin{lstlisting}[language=SMTLib,frame=single]
(assert (str.in_re w (re.from_ecma2020 "[a-z]*|[0-9]*")))
\end{lstlisting}
To enable full support of ECMAScript character escaping, OSTRICH2 also supports
single-quoted strings, in which the normal SMT-LIB character escaping
is disabled and strings are passed unmodified to the
regular expression parser:
\begin{lstlisting}[language=SMTLib,frame=single]
(assert (str.in_re w (re.from_ecma2020 '\x24[0-9]+')))
\end{lstlisting}

One notable extension is the use of capture groups and references in replace and extract functions.
The following example shows the replacement of any substring matching \verb+src='.*'+ with the text appearing between the single quotation marks.
For example
\verb+<img src='image.png'>+ would become
\verb+<img image.png>+.
We use the \verb+re.from_ecma2020+ function for convenience;
direct functions for regular expression operators---such as \verb+re.capture+---are also available.
\begin{lstlisting}[language=SMTLib,frame=single]
(assert (= x
           (str.replace_cg_all
              y
              (re.from_ecma2020 "src='(.*)'")
              (_ re.reference 1))))
\end{lstlisting}
%\todo[inline]{MH: do we need (\_ re.reference 1) or is there a
%  fromECMA alternative? PR: not at the moment, but would be easy to add}

There are several features to note about this example.
The first argument of \verb+str.replace_cg_all+ is the string to be searched.
The second is the regular expression that captures part of the string using the parenthesis notation \verb+(+\ldots\verb+)+.
The final argument is the \emph{replacement pattern} that can be a concatenation of string literals and references to captured text.
In this case, only the contents of the first (and only) capture group are used.

Importantly, when matching \verb+src='(.*)'+, precedence rules must be respected.
The \verb+*+ operator is \emph{greedy} in ECMAScript (and many other regular expression languages), which means that it should match as many characters as possible before the remainder of the match begins.
This can be a cause of errors when developers write regular expressions and a challenge for symbolic execution tools using a string constraint solver~\cite{LMK19}.
For example, if \verb+y+ contained the value \verb+src='a' src='b'+, then there should only be one match of \verb+src='(.*)'+ rather than two.
In the match, the captured value is \verb+a' src='b+,
that is, the longest string surrounded by \verb+'+ symbols.
This semantics is respected by OSTRICH2 through the use of \emph{prioritised} transducers~\cite{BDM14,BM17,popl22}.
In a prioritised transducer, steps of the transduction can be given priority over the alternatives.
When matching against \verb+src='a' src='b'+ there is a choice whether the match of \verb+(.*)+ should stop at the second \verb+'+ or continue.
Because \verb+*+ is greedy, the match of \verb+(.*)+ has priority and
matching can only stop if continuing would fail. When replacing the
pattern
\verb+src='(.*)'+ with \verb+src='(.*?)'+,
using a \emph{lazy} quantifier~\verb+*?+, instead the
shortest string will be matched.

\subsubsection{Automata Representations}

OSTRICH2 also directly supports finite-state automata, which are often
more convenient than regular expressions when integrating an SMT
solver in applications.
This automaton syntax is a bespoke extension specific to OSTRICH2. 
To the best of our knowledge, no other SMT solver supports this format.
The function \verb|re.from_automaton| parses a finite‑state automaton given as a string.
In the example below the automaton has
\begin{itemize}
    \item an initial state \texttt{s0},
    \item an accepting state \texttt{s1},
    \item a transition $\texttt{s0 -> s1 [0,100];}$ (accepting character codes 0–100),
    \item a loop $\texttt{s1 -> s1 [0,65535];}$ (accepting any UTF-16 code unit).
\end{itemize}
Then \verb|str.in_re| can be used to test whether the value of an expression belongs to the language of that automaton.

\begin{lstlisting}[language=SMTLib,frame=single]
(assert (str.in_re x
       (re.from_automaton "automaton value_0 {init s0; s0 -> s1 [0, 100]; s1 ->s1[0,65535]; accepting s1;};")))
\end{lstlisting}

A formal description is given in Appendix \ref{app:automata-grammar}.

\subsubsection{SMT-LIB Standard for Unicode Theory of Strings}

We begin by outlining the basic syntactic elements for our formulas and the key string and regular expression operations as defined in the SMT-LIB Unicode theory.
The fragment below shows the core string and regular expression grammar supported natively by OSTRICH2. 
Functions not explicitly listed here are internally translated or reduced to equivalent formulas in this grammar. 
Operators that are extensions beyond SMT-LIB~2.6 are shown \underline{underlined}.

\noindent\textbf{Formulas:}
\[
\begin{array}{rcl}
	\phi & ::= & \lnot\, \phi \mid \phi \land \phi \mid \phi \lor \phi \mid Atom  \\[1ex]
	Atom & ::= & t_{\mathit{s}} \sim t_{\mathit{s}} \mid t_{\mathit{ar}} \sim t_{\mathit{ar}} \mid t_{\mathit{s}} \in t_{\mathit{re}} \mid \\
	&     & StrPred \mid \underline{\mathcal{T}(t_s,t_s)}\\[1ex]
	StrPred & ::= & \operatorname{prefixof}(t_{\mathit{s}}, t_{\mathit{s}}) \mid \operatorname{suffixof}(t_{\mathit{s}}, t_{\mathit{s}}) \mid \\[1ex]
	&     & \operatorname{contains}(t_{\mathit{s}}, t_{\mathit{s}}) \\[1ex]
	t_{\mathit{s}} & ::= & BaseStr \mid StrPos \mid StrRep \mid \\ &     &\underline{\operatorname{reverse}(t_{\mathit{s}})} \\[1ex]
	BaseStr & ::= & c_{\mathit{str}} \mid x_{\mathit{str}} \mid  \sconcat(t_{\mathit{s}},t_{\mathit{s}})\  \\[1ex]
	StrPos & ::= & \operatorname{at}(t_{\mathit{s}}, t_{\mathit{ar}}) \mid \operatorname{substr}(t_{\mathit{s}}, t_{\mathit{ar}}, t_{\mathit{ar}}) \\[1ex]
	StrRep & ::= & \operatorname{rep}(t_{\mathit{s}}, t_{\mathit{s}}, t_{\mathit{s}}) \mid \operatorname{rep\_all}(t_{\mathit{s}}, t_{\mathit{s}}, t_{\mathit{s}}) \mid \\[1ex]
	& & \operatorname{rep\_re}(t_{\mathit{s}}, t_{\mathit{re}}, t_{\mathit{s}}) \mid \operatorname{rep\_re\_all}(t_{\mathit{s}}, t_{\mathit{re}}, t_{\mathit{s}}) \\[1ex]
	t_{\mathit{ar}} & ::= & c_{\mathit{int}} \mid x_{\mathit{int}} \mid t_{\mathit{ar}} + t_{\mathit{ar}} \mid \lvert t_{\mathit{s}} \rvert \mid \\[1ex]
	&     & \operatorname{indexof}(t_{\mathit{s}}, t_{\mathit{s}}, t_{\mathit{ar}}) \\[1ex]
	t_{\mathit{re}} & ::= & \emptyset \mid \Sigma \mid \Sigma^* \mid \operatorname{toRE}(t_{\mathit{s}}) \mid t_{\mathit{re}} \cdot t_{\mathit{re}} \mid \\[1ex]
	&     & t_{\mathit{re}} \cup t_{\mathit{re}} \mid t_{\mathit{re}} \cap t_{\mathit{re}} \mid t_{\mathit{re}}^*
\end{array}
\]

\section*{Explanation}

\begin{itemize}
	\item \textbf{Formulas (\(\phi\)) and Atoms:}  
	The Boolean formulas \(\phi\) are constructed using the standard logical connectives (negation \(\lnot\), conjunction \(\land\), and disjunction \(\lor\)) along with atomic formulas. The atomic formulas include comparisons on string terms \( t_s \) (via a placeholder relation \(\sim\)), arithmetic terms \( t_{\mathit{ar}} \) (also using \(\sim\)), the membership test \( t_s \in t_{\mathit{re}} \), and the dedicated string predicates defined by \( StrPred \).
	Finally, \underline{$\mathcal{T}(t_s,t_s)$} encodes a transducer.
	The complete grammar for transducers is provided in Appendix~\ref{app:transducer-grammar}.
	
	\item \textbf{String Predicates (\(StrPred\)):}  
	This subclass of atoms is dedicated to predicates that operate specifically on strings. It includes:
	\begin{itemize}
		\item \(\operatorname{prefixof}(t_s, t_s)\): Checks if the first string is a prefix of the second.
		\item \(\operatorname{suffixof}(t_s, t_s)\): Checks if the first string is a suffix of the second.
		\item \(\operatorname{contains}(t_s, t_s)\): Checks if the first string is contained within the second.
	\end{itemize}
	
	\item \textbf{String Terms (\(t_s\)):}  
	The nonterminal \( t_s \) is divided into three subclasses and one string function:
	\begin{itemize}
		\item \textbf{\(BaseStr\):}  
		Represents the basic string values. This includes string constants \(c_s\), string variables \(x_s\), or the concatenation of two string terms using the operator \(\sconcat\).
		
		\item \textbf{\(StrPos\):}  
		Represents string functions that require an integer parameter. These include:
		\begin{itemize}
			\item \(\operatorname{at}(t_s, t_{\mathit{ar}})\): Returns the character (as a string) at a specified position.
			\item \(\operatorname{substr}(t_s, t_{\mathit{ar}}, t_{\mathit{ar}})\): Returns a substring starting at a given position with a specified length.
		\end{itemize}
		
		\item \textbf{\(StrRep\):}  
		Contains the string replacement functions, abbreviated here for conciseness. They include:
		\begin{itemize}
			\item \(\operatorname{rep}(t_s, t_s, t_s)\): Literal replacement.
			\item \(\operatorname{rep\_all}(t_s, t_s, t_s)\): Replace all occurrences (literal).
			\item \(\operatorname{rep\_re}(t_s, t_{\mathit{re}}, t_s)\): Regex-based replacement.
			\item \(\operatorname{rep\_re\_all}(t_s, t_{\mathit{re}}, t_s)\): Replace all occurrences based on a regex.
		\end{itemize}
	\end{itemize}
	
	Finally, the operator \underline{\(\operatorname{reverse}\)} reverses a string.

	\item \textbf{Arithmetic Terms (\(t_{\mathit{ar}}\)):}  
	The arithmetic expressions include integer constants \(c_{\mathit{int}}\), integer variables \(x_{\mathit{int}}\), the sum of two arithmetic expressions \(t_{\mathit{ar}} + t_{\mathit{ar}}\), the length function \(\lvert t_s \rvert\) (which returns the length of a string), and the function \(\operatorname{indexof}(t_s, t_s, t_{\mathit{ar}})\), which returns the index of one string within another as an integer.
	
	\item \textbf{Regular Expressions (\(t_{\mathit{re}}\)):}  
	Regular expressions are defined in a manner consistent with standard mathematical notation:
	\begin{itemize}
		\item \(\emptyset\) denotes the empty language.
		\item \(\Sigma\) is the alphabet, and \(\Sigma^*\) represents its Kleene closure.
		\item \(\operatorname{toRE}(t_s)\) converts a string term into its corresponding regular expression.
		\item The operators \(\cdot\) (concatenation), \(\cup\) (union), \(\cap\) (intersection), and \(^*\) (Kleene star) are used to build more complex regular expressions.
	\end{itemize}
\end{itemize}

	\section{System Architecture}\label{sec:architecture}
	
\begin{figure*}[t]
    \begin{center}
     \begin{tikzpicture}[
     	>=Latex,
     	font=\small,
     	node distance=8mm and 12mm,
     	block/.style={draw, rounded corners, align=center, minimum width=37mm, minimum height=7mm, fill=white},
     	title/.style={align=center},
     	bigbox/.style={draw, rounded corners=6pt, fill=black!3, inner sep=6pt},
     	arr/.style={-Latex, line width=0.7pt},
     	]
     	% Style for big boxes
     	\tikzset{
     		bigboxlabel/.style={
     			font=\bfseries,
     			anchor=south west,
     			xshift=2pt,
     			yshift=-2pt
     		},
     		solverbox/.style={draw, rounded corners, thick, inner sep=4pt, fill=blue!10},
     		rulesbox/.style={draw,  thick, inner sep=4pt, fill=orange!10}
     	}

		% --- Layout with a 3-column matrix ---
		\matrix[matrix of nodes, column sep=16mm, row sep=5mm, nodes={anchor=center}] (M) {
			% Row 1
			|[block] (input)| SMT-LIB Input        & |[block] (inter)|    Automata Intersection      & |[block] (sdb)| String Database\\
			% Row 2
			|[block] (pre)  | String Preprocessor  & |[block] (fwd) | FWD Propagation & |[block] (adb)| Automata Database \\
			% Row 3 (spacer in col 1, real content in col 2, real content in col 3)
			& |[block] (bwd)  | BWD Propagation       &   \\
			% Row 4
			|[block] (core) | SMT Core             & |[block] (nielsen)  | Nielsen Splitting       & |[block] (brics)| BRICS Automata Library \\
			% Row 5
			|[block] (lia)  | LIA/ADT Solver       & |[title] (etc)| etc.   & \\
		};

     	% Rename the top-left title node for arrows
		\begin{pgfonlayer}{background}
			\node[solverbox, fit=(core)(lia)] (princessbox) {};
			\node[bigboxlabel] at (princessbox.north west) {Princess};
		\end{pgfonlayer}
		
				% Big box around middle column
		\begin{pgfonlayer}{background}
			\node[rulesbox, fit=(inter)(fwd)(bwd)(nielsen)(etc)(etc.south|-lia.south)] (rulesbox) {};
			\node[bigboxlabel] at (rulesbox.north west) {RCP/CEA Rules};
		\end{pgfonlayer}

		\draw[arr] (input.south) -- (pre.north);
		\draw[arr] (pre.south) -- (princessbox.north);
		\draw[arr] (core.south) -- (lia.north);
		
		\draw[arr,<->]
		([yshift=-18.5mm]princessbox.east |- rulesbox.center) --
		([yshift=-18.5mm]rulesbox.west   |- rulesbox.center);

		% String Database <-> Proof Rules  (use y from sdb)
		\draw[arr,<->]
		(sdb.west) --
		(rulesbox.east |- sdb.center);
		
		% Automata Database <-> Proof Rules (use y from adb)
		\draw[arr,<->]
		(adb.west) --
		(rulesbox.east |- adb.center);

		% Automata Database <-> Brics Automata Library
		\draw[arr,<->]
		(adb.south) -- (brics.north);

     \end{tikzpicture}
     
    \end{center}
    \caption{Overall architecture of OSTRICH2: the SMT-LIB input is handled by our string preprocessor and Princess SMT core (with LIA and ADT-Str), while the RCP and CE-Str solvers consist of inference rules that are repeatedly applied using  string and automata databases.}
    \label{fig:architecture}
\end{figure*}
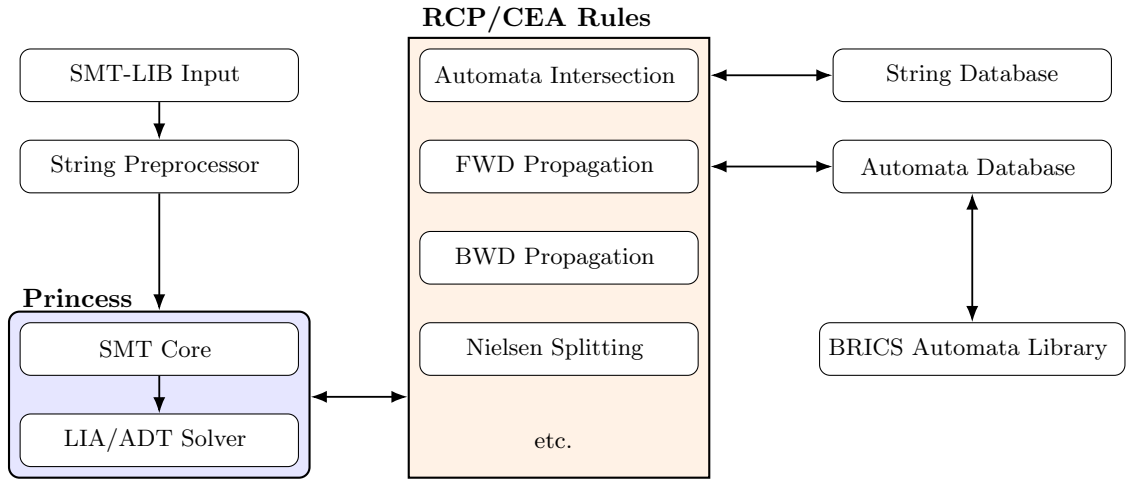

The overall architecture of OSTRICH2 is depicted schematically in Figure~\ref{fig:architecture}.
OSTRICH2 runs the three solvers (ADT-Str, RCP, and CE-Str) sequentially in a time-sliced configuration. Each solver receives the same input problem and runs independently for a fixed share of the total timeout (e.g. 20 seconds each for a 60-second limit). To reduce repeated work, OSTRICH2 performs common subexpression sharing for selected functions while representing all other expressions separately in tree form. Once started, no further information is shared between solvers as there is currently no established methodology in the literature for aggressive or global sharing of string expressions.

At its core, OSTRICH2 builds on the SMT solver Princess~\cite{princess}, which provides the logical reasoning framework and support for theories such as linear integer arithmetic.
OSTRICH2 provides pre-processing and support for the quantifier-free string theories \verb+QF_S+ (pure strings) and \verb+QF_SLIA+ (strings with linear integer arithmetic).
Princess also provides support for algebraic data-types (ADTs), on top
of which the ADT-Str string solver is implemented.

The second and third solver are the Regular Constraint Propagation
(RCP) and Cost-Enriched String (CE-Str) engine, respectively.
These share similar architectures.
They begin with a shared string preprocessor, which processes the SMT-LIB input and forwards it to the SMT Core of Princess.
Details of the string preprocessing are given in the next section.
The SMT Core coordinates with the LIA Solver for linear integer arithmetic and with a solver engine for string-specific reasoning.

Both solvers uses automata- and tranducer-based representations of sets of strings and supported string functions.
A main loop applies proof rules like Automata Intersection, Forward (FWD) and Backward (BWD) Propagation, Nielsen Splitting, and other string-related inference rules.
Details of the rules are given in the next section.
A string database and an automata database are used for efficiently storing and retrieving string and automata data.
The automata database is based on the BRICS Automata Library~\cite{brics}, which provides efficient automaton operations.

For the purposes of presentation, we abstract the constraints received from
the SMT core into the following \emph{normal form}:
\begin{align*}
	S ~~::=~~ & A \;|\; \neg A \;|\; S \land S \\
	A ~~::=~~ &
	f(x_1, \ldots, x_n)
	\;|\; x = g(x_1, \ldots, x_n)
	\;|\; \\ & x \in L
	\;|\; x = \sum_i d_i x_i \ .
\end{align*}
In this notation, \(x, x_1, \ldots, x_n\) are variables of type string or integer, and each \(d_i\) is an integer constant. The symbol \(f\) denotes a string predicate such as \(\mathsf{prefixof}(x_1, x_2)\) or \(\mathsf{suffixof}(x_1, x_2)\). The symbol \(g\) denotes a string function that takes \(x_1, \ldots, x_n\) as inputs and produces the string \(x\) as output, for example \(x = \mathsf{concat}(x_1, x_2)\). The notation \(x \in L\) expresses that the value of \(x\) belongs to a regular language \(L\), which is typically given by a regular expression in SMT-LIB syntax. 
Finally, the form \(x = \sum_i d_i x_i\) represents a linear integer constraint, most commonly arising from length constraints, where the \(x_i\) are integer variables and the \(d_i\) are constant coefficients. 
We sometimes write $|t|$ to denote the length of the word represented by the term $t$.
For example, the constraint
\begin{lstlisting}[language=SMTLib,frame=single]
(assert (and (= (str.++ x y) (str.++ y x))
             (str.in_re x (re.from_ecma2020 "a*ba*"))
             (str.in_re y (re.from_ecma2020 "a*ca*"))))
\end{lstlisting}
has the following normal form using an additional variable $z$
\[
    z = \sconcat(x,y) \wedge z = \sconcat(y,x)
        \wedge x \in a^*ba^* \wedge y \in a^*ca^* \ .
\]

%ecma script 2020 -> remove

In the sections below, we give an overview of the three main solver engines.
These are: ADT-Str (algebraic data-types), RCP (regular constraint propagation), and CE-Str (cost-enriched strings).

\subsection{ADT-Str: List-Based Solver}

The ADT-Str solver builds on the decision procedure for algebraic data-types (ADTs) with size catamorphism implemented in Princess~\cite{DBLP:conf/synasc/HojjatR17}.
Algebraic data-types are used to represent strings using the standard encoding of lists with \texttt{nil} and \texttt{cons} constructors.
The length of a string is computed using the built-in \texttt{size} function provided by the ADT solver, mapping every constructor term to the number of constructor occurrences.
Other SMT-LIB functions on strings, for instance substring, concatenation, etc., are in ADT-Str encoded using uninterpreted functions and axioms capturing the recursive
definition of the string functions.
Regular expression matching is implemented using Brzozowski derivatives~\cite{Brzozowski}.

ADT-Str is  useful, in particular, for computing solutions of string
constraints that are outside of the fragments for which the other
solvers are complete.
ADT-Str can easily handle certain functions that are hard for our propagation algorithms.
Those functions include, among others, string-to-integer conversion, and functions like substring and indexof that calculate with integer offsets.

\subsection{RCP: Regular Constraint Propagation}

\emph{Regular Constraint Propagation (RCP)} is the newest algorithm that has been implemented in OSTRICH2, based on a subset of proof rules in our paper~\cite{popl22}.
The main goal of RCP is to \emph{prove unsatisfiability of the input constraint}.
The algorithm handles string functions like concatenation, replace, replaceall, and regular constraints.
Other string functions (e.g.\ reverse, one-way and two-way transducers) are also permitted.
These functions permit either exact pre-image or exact post-image computation of regular constraints or both.
In general, these images need not be exact, but must at least overapproximate the true pre/post image.

The main idea behind RCP is to propagate regular constraints of the form
$x \in L$ to other string variables through forward and backward propagations using the RCP inference rule described in the next section.
\subsubsection*{ECMAScript Regular Expressions}
All three solvers in OSTRICH2 support the SMT-LIB regular expression operators. 
Full support for ECMAScript regular expression features, including look-arounds, capture groups, and greedy/lazy quantifiers, is implemented in the RCP solver, following the approach in earlier work~\cite{popl22,blackostrich}. 
Back-references are supported only in the replacement string of \texttt{replace} and \texttt{replace\_all}, not in the matching expression.
This implementation uses prioritised transducers to preserve ECMAScript semantics and an intermediate translation to alternating two-way automata for look-arounds.
We refer to~\cite{popl22,blackostrich} for full technical details.

\subsection{CE-Str: Cost-Enriched String Engine}

The main algorithm of CE-Str (based on \cite{atva2020}) applies backward regular constraint propagation.
It uses cost-enriched finite automata (CEFAs) instead of standard finite automata to represent sets of words.
These automata are able to capture numerical information about accepted strings, such as the string length, the number of characters appearing before a transition is fired, and so on.
This allows length constraints on strings and functions like indexof to be directly supported.

In addition, CE-Str can handle counting operators more efficiently by leveraging the numerical information encoded in CEFAs. For instance, the regex $a^{\{1,1000\}}$, which accepts strings consisting of 1 to 1000 repetitions of the character $a$, can be represented as a CEFA with just one state and one transition—compared to the 1000 states and 1000 transitions required by a standard finite automaton.

CE-Str supports a wide range of string functions, including concatenation, replace, replaceall, substring, indexof and length. It provides completeness guarantees for the straight-line fragment of these functions. CE-Str complements RCP and ADT-Str. It performs well on string constraints involving integer type but struggles when the constraints are outside its decidable fragment.

	\section{String Theory Algorithms}\label{sec:string-theory-algorithms}
	% !TeX root = main.tex

In this section we describe the core inference rules and algorithmic components that OSTRICH2 applies to solve string constraints.
We first present the preprocessing steps shared by the RCP and CE-Str solvers.
We then present the \emph{inference rules} applied in the main loops of RCP and CE-Str.

\subsection{Preprocessing}

Before applying the core inference rules, OSTRICH2 performs a series of preprocessing steps to enrich the constraint set with auxiliary information and simplify trivial cases.
The simplification rules are given below.
First we describe the simplification rules used by both RCP and CE-Str, then we describe the additional rules used by RCP only.
%\paragraph{Preprocessor vs.\ Reducer}
%OSTRICH actually performs two flavors of “pre-solving” simplifications.  The standalone \emph{Preprocessor} (in \texttt{OstrichPreprocessor.scala}) runs once on the raw SMT-LIB input, under a global context: it desugars high-level constructs (regex literals, transducers), introduces fresh variables, applies context-sensitive rewrites (e.g.\ inlining of definitions, constant folding) and emits a flat conjunction of normalized constraints.  By contrast, the \emph{Reducer} lives inside the string‐solver loop—operating on each branch’s current conjunction in CNF—and applies stronger, in-context simplifications that only make sense once some constraints are already solved or partially evaluated.  These include discharging newly trivial predicates, merging overlapping automata constraints, eliminating solved variables, and propagating length/character bounds discovered during proof construction.  Splitting these two stages lets us handle the full SMT-LIB grammar up-front, while still enjoying lightweight, incremental reductions during solving.

\subsubsection{Common Preprocessing}

\paragraph*{Prefix/Contains/Suffix Simplification}
We apply the following rewrites when one argument is concrete:
\begin{itemize}
	\item $\sprefixof(s,t)$ / $\ssuffixof(s,t)$ with concrete $s$:
	replace by
	\[
	t \;\in\; L(\texttt{s.*}) \quad/\quad t \;\in\; L(\texttt{.*s}).
	\]
	\item $\scontains(t,s)$ with concrete $s$:
	replace by
	\[
	t \;\in\; L(\texttt{.*s.*}).
	\]
\end{itemize}
In \emph{positive} contexts we instead encode:
\[
    \begin{array}{rcl}
        \sprefixof(s,t)&\rightsquigarrow&t = \sconcat(s, u) \\
        \ssuffixof(s,t)&\rightsquigarrow&t = \sconcat(u, s)
    \end{array}
\]
for fresh $u$.
Finally, $\sprefixof(t,t)$, $\scontains(t,t)$, $\ssuffixof(t,t)$
are \emph{trivially true} and replaced by the Boolean constant $\mathsf{true}$.

\paragraph*{Common Sub-Expression Elimination}

While OSTRICH2 does not apply aggressive sharing of terms and expressions, it is useful to merge repeated occurrences of certain operators that are relatively expensive to handle, including expressions~$\sindexof(x,y,i)$. For expressions~$e$ of this kind that occur multiple times in a formula, OSTRICH2 introduces a fresh variable~$k$ and adds the equation~$k = e$.
All occurrences of $e$ in the formula are then rewritten as $k$.
This rewriting ensures that the solver only sees one copy of each complex sub-expression and speeds up subsequent reasoning.

%To avoid repeatedly rewriting the same complex subexpression (e.g.\ multiple occurrences of $\sindexof(x,y,i)$), the preprocessor replaces each closed occurrence by a fresh constant $k$ and records the definition $k = \sindexof(x,y,i)$.  Internally this is encoded as
%\[
%\forall k.\;\bigl(k = \sindexof(x,y,i)\bigr)\;\Longrightarrow\;\phi\bigl[k/\sindexof(x,y,i)\bigr],
%\]
%so that $k$ is scoped like a local `let`.  Although this formula contains a universal quantifier, it is only used to simulate local binding of $k$.  Before the solver sees the constraints, we substitute each definition $k = e$ back into the body and drop the quantifier, preserving a purely quantifier‐free conjunction of equalities.

%    \todo[inline]{MH: given the supported logic is quantifier-free, what does this mean?}
 %       \todo[inline]{OM: I wrote 2 versions of text to answer this, detailed and simplified version.}
%        \todo[inline]{MH: the both look quite different! The first means the solver gets constraints with just k in them, the second means the solver is handling the original constraint, but in a "compressed" form that is expanded on the fly. Which one is it?}

\subsubsection{RCP Preprocessing}

The next simplifications are currently used by RCP only.

\paragraph{Length and Character-Count Approximation}
Infers approximate bounds on string lengths and character counts for complex string functions.
\begin{itemize}
    \item \emph{Derived length constraints:} derive length constraints and character count approximations from word equations and other string functions.
    E.g.\ $|x| = |y| + |z|$ is added for $x = \sconcat(y, z)$.
    \item \emph{Automaton‐transition analysis:} inspect each transition in an automaton $A$ and, for any character $a$ not appearing on any transition, add the char count constraint $|x_a| = 0$
    for every $x$ constrained by $A$.
    \item \emph{Index‐of range constraints:} for each occurrence of $\sindexof(x,y,i)$ (which returns the first index of $y$ in $x$ after position $i$), add implied constraints on the result.
    E.g.,
    $-1 \leq \sindexof(x, y, i) \leq \lvert x\rvert$
    captures that the result must be either be $-1$ or in the interval $[0,\lvert x\rvert]$.
\end{itemize}

\paragraph{Index‐of/Substr/At Rewriting}
Translate $\ssubstr$, $\sat$, and $\sindexof$ into equivalent combinations of string concatenation, length constraints, and regex‐membership tests by introducing fresh variables and constraints to encode the positional semantics.

%\[
%str.substr(s,i,n)
%\;=\;
%\begin{cases}
%	\varepsilon, &
%	i<0\;\lor\;n\le0\;\lor\;i\ge |s|, \\[4pt]
%	r, &
%	\begin{aligned}[t]
%		\exists\,p,r,q.\;s=p\,r\,q\;\wedge\;|p|=i\;\wedge\\
%		|r|=n \land n\le |s|-i
%	\end{aligned}
%	\\[4pt]
%	r, &
%	\begin{aligned}[t]
%		&\exists\,p,r,q.\;s=p\,r\,q\;\wedge\;|p|=i\;\wedge\\ &|r|=|s|-i \land		i<|s|\;\land\;i+n>|s|
%	\end{aligned}
%\end{cases}
%\]
We illustrate one such rewrite for $r = \ssubstr(s,i,n)$, which asserts that $r$ contains the longest contiguous substring of $s$ of length at most $n$ starting at position $i$.
It can be split into three cases.

First we introduce fresh string variables $p,r,q$ with
\[
s = \sconcat(p, r, q) \land |p| = i,
\]
and then distinguish:

\begin{itemize}
	\item If $i<0$, $n\le0$, or $i\ge|s|$, then $r = \varepsilon$.

	\item If $0 \le i < |s|$ and $i + n \le |s|$, then $|r| = n$, so $r$ is exactly the length-$n$ slice of $s$ starting at position $i$.

	\item If $0 \le i < |s|$ but $i + n > |s|$, then $|r| = |s| - i$, so $r$ is the substring from position $i$ to the end of $s$.
\end{itemize}

\subsection{Inprocessing Rules}
\label{sec:inprocessing}

There is also a set of lightweight simplification rules that are applied to
expressions during proving. Such rules are applied in the local
context of a proof goal and are often able to significantly simplify
expressions, for instance by evaluating function applications with
known arguments or discovering obvious contractions. During
inprocessing, equations and assignments of values to variables are
propagated to other constraints. There are also
certain preprocessing rules that are applied again during
inprocessing; for instance, $\neg\,\ssuffixof(s,t)$ can be turned into
a regular expression constraint as soon as one of the arguments~$s, t$
is a known string, but has to be kept unchanged before that.

\iffalse
\begin{itemize}
	\item eliminate newly obvious contradictions or tautologies (e.g.\ if an automaton constraint becomes empty, close the branch)
	\item re-apply any preprocessing-style rewrites that depend on polarity (for example, rewrite $\neg\,\ssuffixof(s,t)$ to membership tests and disequalities)
	\item incorporate any new equalities or bounds discovered by other inference rules such as a fresh assignment from lazy enumeration or from the cut rule.
\end{itemize}

Not all of these rewrites can be performed in the up-front preprocessor, since at that point we don’t yet know which literals will be negated or which variables will become concrete during solving.

\todo[inline]{MH: the first bullet is reasonable, but the second sounds like ``do other things like the list above'', which is somewhat mysterious.}
\fi

\subsection{Inference Rules}
\label{sec:inference-rules}

The main loops of the RCP and CE-Str repeatedly apply proof rules until a branch is closed or a model is found. We discuss the principal rules in this section. The additional rules omitted here that which cover both string-specific reasoning and SMT-core inferences are described in previous work~\cite{popl22}, and together with the rules below constitute a complete proof system.

The RCP algorithm assigns priorities to each possible rule application based on an estimation of the workload (e.g.\ the size of the involved automata) and selects the rule application with the highest priority first. Fairness is ensured by penalizing newer rule applications, and therefore preferring rule applications that have resided in the waiting queue the longest.
%
%In RCP we distinguish two independent queues—one for forwards propagation and one for backwards propagation.  When the SMT core decides to do a forward step, RCP scores all pending forward‐propagation tasks and fires the highest‐scoring one; likewise, a backward step picks from the backward queue.
Priorities are computed as the weighted sum of several criteria:
\begin{itemize}
	\item \emph{Concrete‐argument:} any rule whose input or output language is a ground string is given high priority.
	\item \emph{Information‐gain penalty (backward only):} rules whose input automata are universal (i.e.\ accept all words, e.g.\ $x\in\Sigma^*$) yield little new information in backward propagation and given low priority.
	\item \emph{Exactness adjustment (forward only):} forward rules for functions without an exact post-image (e.g.\ replaceall with symbolic patterns) are given low priority.
	\item \emph{Cost‐based penalty:} a weight proportional to the combined size of the input and result automata is subtracted.
\end{itemize}

%We fully saturate forward‐ and backward‐propagation (including length abstraction) before applying any other inference rules; whenever new information arrives for a string variable, we re-enter this saturation phase.

%\todo[inline]{MH: Philipp can you say something proper here?}
%\todo[inline]{PR: actually, everything works with priorities, so it is never the case that we first fully apply one rule before another rule. It would be difficult to ensure completeness using such strict priorities.}

Differences between the two solvers in their application of the rules is included in the descriptions.
The proof rules may cause branching in the proof search.
If a satisfying assignment is found on a branch, it witnesses that the constraint is satisfiable.
The Close rule detects unsatisfiable constraints.
If all branches are closed, the constraint is unsatisfiable.

\paragraph{Regex‐to‐Automata}
OSTRICH2 constructs an automaton $A$ for each regular membership constraint in $x \in L$.
The automaton is stored in the automaton database, which detects equivalent automata to avoid duplication.
In the remaining rules, we make the automaton explicit by writing $x \in L(A)$.
The CE-Str solver creates cost-enriched automata.
Initially these automata do not track any costs.
Cost tracking is introduced during the RCP rule below.

\paragraph{BreakCyclicEquations}
OSTRICH2 detects strongly connected components in the variable‐dependency graph induced by concatenation equations of the form
\[
x = \sconcat(y,z)
\quad\text{and}\quad
y = \sconcat(a,x).
\]
For each cycle it removes one equation and, for every remaining equation $\,v = \sconcat(u,w)$ in that cycle, adds a trivial emptiness constraint on one argument (e.g.\ $w = \varepsilon$ or $u = \varepsilon$) to break the dependency.

For example, from
\[
x = \sconcat(y,z)
\quad\land\quad
y = \sconcat(a,x)
\]
we might remove the first equation and introduce
\[
y = \sconcat(a,x),
\quad
z = \varepsilon,
\quad
a = \varepsilon,
\]
yielding an acyclic set of equations.
This transformation is \emph{sound}, because in any finite model of the original cyclic equations at least one concatenation argument must be empty, and \emph{complete}, since any solution of the resulting acyclic system (with the added emptiness constraints) automatically satisfies the dropped equations.

\paragraph{Equation Decomposition}
OSTRICH2 uses two simple heuristics to simplify word equations.  In what follows, $z,x_1,x_2,y_1,y_2$ are arbitrary string terms (each may be a variable or a constant).
First, when one side is a constant string $w$ and the other is a concatenation
\[
w = \sconcat(x_1,x_2),
\]
in which the length of $x_1$ is known,
OSTRICH2 matches the first $|x_1|$ characters of $w$ with $x_1$ and the remaining characters with $x_2$. Second, if the same variable appears in two concatenations,
\[
z = \sconcat(x_1,x_2)
\quad\text{and}\quad
z = \sconcat(y_1,y_2),
\]
then whenever $|x_1| = |y_1|$ we infer $x_1 = y_1$ and $x_2 = y_2$. These heuristics often resolve equations without full case splitting.

\paragraph{Close}
If for some variable $x$ we have constraints
\[
    x \in L(A_1) \land \dots \land x \in L(A_k)
\]
and $\bigcap_{i=1}^k L(A_i)=\emptyset$, OSTRICH2 derives a contradiction and close the branch.

\paragraph{Intersection}
When the \texttt{+eager} flag is on, OSTRICH2 maintains at most one automaton per variable by replacing $x \in L(A) \land x \in L(B)$ with $x \in L(A \cap B)$.

\paragraph{LengthAbstraction}
From length inequalities (e.g.\ $|x|\le|y|+5$), derive lower/upper bounds on $|x|$, then assert $x \in L(A)$ where $A$ accepts any word within the derived length bounds.

\paragraph{RCP (Regular Constraint Propagation)}
Propagate $x_1 \in L_1$, \ldots, $x_n \in L_n$ forwards through string functions $x=f(x_1,\dots,x_n)$, or $x \in L$ backwards through $f$.

For example, on
$z = concat(x,y) \wedge z = concat(y,x) \wedge x \in a^*ba^* \wedge y \in a^*ca^*$
we can propagate $x \in a^*ba^*$ and $y \in a^*ca^*$ forwards (from input to output) through the string concatenation function $concat$ in $z = concat(x, y)$ to obtain $z \in a^*ba^*a^*ca^*$.
Similarly, we can propagate forwards through $z = concat(y, x)$ and derive $z \in a^*ca^*a^*ba^*$.
Since there are no words matching both $a^*ba^*a^*ca^*$ and $a^*ca^*a^*ba^*$ we can conclude---using the Close rule---that the constraint is unsatisfiable.

Backwards propagation may result in branching.
For example, if $x = \sconcat(y, z)$ and $x = ab$, then either $y = ab \land z = \varepsilon$ or $y = a \land z = b$ or $y = \varepsilon \land z = ab$.

When the variable constraints are given by standard finite automata, a rich selection of string functions support exact forwards and backwards propagation.
These include $\sconcat$, $\sreverse$, and $\sreplace$ and $\sreplaceall$, where the replacement pattern can include string variables or references to capture groups in the search pattern.
In general, any function can be supported using over-approximations of the pre- and post-images, but without completeness guarantees.

The CE-Str solver only applies backwards propagation and represents pre-images using cost-enriched automata.
This allows functions such as $\sindexof$ and $\slength$ to be supported (by introducing costs to the automata), but restricts, for example, which versions of $\sreplace$ and $\sreplaceall$ can be analysed precisely.

\paragraph{NielsenSplitter}
OSTRICH2 invokes Nielsen’s transformation \cite{nielsen1,nielsen2} on any pair of equations in our normal form
\[
z = \sconcat(x_1,x_2)
\quad\text{and}\quad
z = \sconcat(y_1,y_2),
\]
where each of $z,x_i,y_i$ may be a variable or a constant.  We
introduce a fresh string variable $w$ and split the proof into two
\emph{prefix‐alignment} cases, guided by current length information in
a similar way as was done in Norn~\cite{DBLP:conf/cav/AbdullaACHRRS15}.
That is
\[
	|x_1|\ge|y_1|, \quad
	x_1 = \sconcat(y_1,\,w), \quad
	\sconcat(w,\,x_2) = y_2
\]
or
\[
	|y_1|\ge|x_1|, \quad
	y_1 = \sconcat(x_1,\,w), \quad
	\sconcat(w,\,y_2) = x_2.
\]

In each branch we align the longer prefix against the shorter one, adding both the corresponding concatenation equalities and the derived length equality (e.g.\ $|x_1|=|y_1|+|w|$ in the first case).

\paragraph{String-Integer-Conversions}
\iffalse \todo[inline]{ MH: what is the next paragraph referring to?
  Can we stick to the normal form we said we expect rather than $|t|$
  for an arbitrary term (which is not the normal form)?  Is this
  paragraph trying to say that whenever it finds a term $|t|$ it
  starts adding $|t| < 2^b$ constraints and checking satisfiability?
  In normal form this would be $x = f(x_1, \ldots, x_n)$, then i think
  it's ok to say we assert $|x| < 2^b$ (as the normal form is a bit
  long and saying we assert it is ok since we assert an equivalent
  expression).  } \fi For handling expressions $\sstrtoint(s) = n$ or
$\sinttostr(n) = s$, OSTRICH2 systematically explores the possible
values of $n$. As soon as $n$ has a concrete value, inprocessing rules
(Section~\ref{sec:inprocessing}) apply that replace the function
application with an equation or a regular expression constraint
describing the possible values of $s$. To make this exploration
perform well in practical cases, the LIA solver utilized in OSTRICH2
applies interval constraint propagation to narrow down the range of
possible values of $n$ as much as
possible~\cite{DBLP:journals/fmsd/BackemanRZ21}. The interval for $n$
is then sub-divided, until eventually only one possible value for $n$
remains. This search will not terminate in general, since the domain
of $n$ can be infinite, but it tends to derive solutions or
contradictions quickly in many practical cases.

\iffalse
OSTRICH2 avoids an up-front infinite split over all integers by interleaving coarse bounding and on-demand guessing.  First it checks whether $|t| < 2^b$ for increasing $b$, so that eventually $t$ is known to lie in the finite interval $\bigl[-2^b+1,\,2^b-1\bigr]$.
Here $t$ denotes any non-constant integer term produced by one of the built-in functions $\sstrtoint(s)$, $\sinttostr(n)$, or as the index $k$ in $\sindexof(s,p,k)$.  Within that interval, it then guesses candidate values in the simple order
$0,\,1,\,-1,\,2,\,-2,\dots$
and for each guess $k$ adds the case $t = k$ to the current goal.
If this branch immediately contradicts other constraints (for instance $k = \sstrtoint(``42'')$ fails until $k=42$).
The process repeats until a satisfying assignment is found or all $k$ in the interval have been tried.
\fi

\paragraph{Index Computation}

When handling $\sindexof(s,p,k)$ on two concrete strings $s,p$,
OSTRICH2 systematically explores possible values of $k$, taking into
account the fact that any value~$k$ not satisfying
$ 0 \le k \le |s| - |p| $ will necessarily lead to the result~$-1$.
In the ``in-range'' branch OSTRICH2 applies interval constraint
propagation to determine the possible values of $k$, sub-dividing the
interval of values until a concrete value~$k$ has been chosen. As soon
as all arguments of $\sindexof$ have concrete values, inprocessing
rules (Section~\ref{sec:inprocessing}) can evaluate the function.

\iffalse
    it again guesses
$k = 0,\,1,\,2,\dots$ pruning each value if $p$ does not occur at that position, and stops as soon as a valid match is found (e.g.\ $\sindexof(\text{"banana"},\text{"na"},k)$ accepts $k=2$). This ensure that solutions to those functions that evaluate to integer are eventually found.

\todo[inline]{%
    MH: needs clarification.
    Is it that when this rule is fired, it picks a possible value for the function result and creates a branch where the result is the picked value and one where it is not?
    What are the length constraints generated?
    For $\sstrtoint$ is it the length of the string encoding of the guessed value?
}

\todo[inline]{%
	OM: I believe that is how it works inside. Philipp probably knows better. In practice $str.to\_int("42")$ should not make it to this point but is caught in the reducer, but what can happen is that we turn $str.to\_int(x)$ to multiple branches, i.e., $str.to\_int(x) = 3$ etc. and then those are checked/rewritten in the reducer.
}
\todo[inline]{MH: Philipp -- can you check?}
\fi

\paragraph{Cut Rule}
When all other rules have been exhausted, OSTRICH2 applies cuts to
introduce a candidate solution for a string variable $x$.  For this, OSTRICH2 collects every regular‐language membership constraint $x \in L_i$
together with any length bounds $\ell\le|x|\le u$.  From the intersection of the $L_i$ (and respecting $\ell,u$), OSTRICH2 extracts a single accepted word $w$ via a standard automaton search.  We then split on the two exhaustive cases
\[
x = w
\quad\text{vs.}\quad
x \notin \{w\}.
\]

This rule is always sound because any satisfying assignment for $x$ either equals the chosen $w$ or lies outside $\{w\}$ and guarantees that each string variable will eventually be grounded to a concrete value (or shown impossible).

	\section{Completeness Results}\label{sec:completeness}
	% !TeX root = main.tex

The RCP solver is (in principle) complete for %orderable fragment~\cite{???}, %which subsumes
the straight-line fragment~\cite{ostrich} and chain-free fragment~\cite{chain19} of string constraints,
in the sense that the underlying proof system can show unsatisfiability of unsatisfiable instances from these fragments,
when applying the rules in an appropriate (easy) order.
However, the usage of priorities means the rules may be applied in a different order, causing the proof to fail.
In practice, failures for these fragments are rare.

In the straight-line fragment we require that the constraints are in normal form as in Section~\ref{sec:inference-rules},
and variables can be ordered $x_1, \ldots, x_n$
such that for $x_i$ there is exactly one equational constraint $x_i = f_i(x_1, \ldots, x_{i-1})$,
where $f_{i}$ is a string function such that the pre-image of a regular set is regular.
The definition of the chain-free fragment is a bit involved, in essence it relaxes the assumption on the form of the equation and the number of equations,
but still requires that there are no ``cycles'' of dependencies.
For the straight-line fragment, we can propagate backwards and the instance is unsatisfiable if and only if all branches are closed using Close rule.
For the chain-free fragment~\cite{chain19} we can adapt the original proof (which works for a different solver) to the rules used in OSTRICH2.

The CE-Str solver also provides a completeness guarantee for a version of the straight-line fragment that includes string constraints and linear integer arithmetic~\cite{atva2020}.
The supported string functions are those where the pre-image of cost-enriched automata constraints can be expressed using cost-enriched automata.
This includes $\sindexof$ and $\slength$ as well as some versions of $\sreplace$ and $\sreplaceall$.

%The orderable fragment is defined procedurally defined: we assume that the constraints are in normal form as in Section~\ref{sec:inference-rules}.
%In each round we mark variables that occur once (in all constraints) then, if there is a constraint $x = f(\overline y)$ such that
%each variable in $\overline y$ is marked and an image of regular set by $f$ is regular (i.e.\ for each regualr $L$ the $f(L)$ is also regular) then
%we remove the constrain (and continue);
%similarly, if $x$ is marked and the pre-image of $f$ of each regular set is a recognizable relation (i.e.\ it is of a form $\bigcup^k_{i=1}
%L^i_1\times \dots \times L^i_n$ for some regular sets $\{L_j^i\}$)
%then we remove the constraint.
%In particular, this algorithm is effective and yields an order in which to apply the \texttt{[Fwd-Prop]} and \texttt{[Bwd-Prop]} rules.

	\section{Extensibility}\label{sec:extensibility}
	% !TeX root = main.tex

An important feature of OSTRICH2 is the possibility to extend the
solver with minimal effort.
OSTRICH2’s \texttt{PreOp} trait lets users add new string functions by
overriding three core methods. This can be done in either Scala (the
main language used to implement OSTRICH2) or in Java.
Below we show each method in turn,
together with the \texttt{ReversePreOp} implementation\footnote{This
  snippet is abbreviated for clarity. The full implementation in
  \texttt{ReversePreOp.scala} includes additional automaton‐type
  matching and error handling.}.

First, the method \texttt{apply} computes the \emph{pre-image} of a
regular language under $f$, which is the core operation needed for
backward propagation in RCP.  The method is passed any existing automata constraints on the $r$ arguments (which may be ignored) and the automaton $\mathcal{A}$ representing the result language.  The method returns an iterator over tuples of automata whose images under $f$ lie inside $\mathcal{A}$, plus (optionally) the subset of the input constraints actually used.

\begin{lstlisting}[language=Scala]
override def apply(
	argCs: Seq[Seq[Automaton]],
	resultC: Automaton
) : (Iterator[Seq[Automaton]], Seq[Seq[Automaton]]) = {
	(Iterator(Seq(ReverseAutomaton(resultC))), List())
}
\end{lstlisting}

Second, the method \texttt{eval} performs \emph{concrete} evaluation on ground strings: when all arguments are known, it simply returns the result of applying $f$.

\begin{lstlisting}[language=Scala]
override def eval(
	args: Seq[Seq[Int]]
): Option[Seq[Int]] = Some(args.head.reverse)
\end{lstlisting}

Third, \texttt{forwardApprox} computes a (sound) \emph{post-image} of input languages under $f$, used for forward propagation.  While returning the universal automaton is always correct, a tighter approximation greatly improves performance.  For \texttt{str.reverse}, one can exactly compute this by intersecting any input automata and then reversing the resulting automaton.

\begin{lstlisting}[language=Scala]
override def forwardApprox(
	argCs: Seq[Seq[Automaton]]
): Automaton = {
	val prod = ProductAutomaton(argCs)
	ReverseAutomaton(prod)
}
\end{lstlisting}
%
%Putting it all together:
%
%\begin{lstlisting}[language=Scala]
%package ostrich.preop
%
%object ReversePreOp extends PreOp {
%	override def name: String  = "str.reverse"
%	override def arity: Int     = 1
%
%	// 1) Pre-image
%	override def apply(
%		argCs: Seq[Seq[Automaton]],
%		resultC: Automaton
%	): (Iterator[Seq[Automaton]], Seq[Seq[Automaton]]) = {
%		(Iterator(Seq(ReverseAutomaton(resultC))), List())
%	}
%
%	// 2) Concrete evaluation
%	override def eval(
%		args: Seq[Seq[Int]]
%	): Option[Seq[Int]] = Some(args.head.reverse)
%
%	// 3) Post-image
%	override def forwardApprox(
%		argCs: Seq[Seq[Automaton]]
%	): Automaton = {
%		val prod = ProductAutomaton(argCs)
%		ReverseAutomaton(prod)
%	}
%}
%\end{lstlisting}

Finally, the new string function has to be registered in the core
string theory class of OSTRICH2.
In the file \texttt{OstrichStringTheory.scala}, we register \texttt{str.reverse} by first declaring its SMT-side function symbol:

\begin{lstlisting}[language=Scala]
val str_reverse = MonoSortedIFunction(
	"str.reverse",
	List(StringSort), StringSort, true, false
)
\end{lstlisting}

We then add it to the \texttt{extraStringFunctions} list, pairing the name, the \texttt{IFunction}, our \texttt{ReversePreOp} implementation, and the simple lambdas that pick out the input and output terms:

\begin{lstlisting}[language=Scala]
val extraStringFunctions = List(
	("str.reverse", str_reverse, ReversePreOp,a =>
		 List(a(0)),a => a(1)))
\end{lstlisting}

With these two lines in place, any occurrence of \texttt{str.reverse} in the SMT-LIB script is recognized by OSTRICH2 and automatically dispatched to our \texttt{ReversePreOp} for pre-image computation, concrete evaluation, and forward-approximation.

	\section{Experiments}\label{sec:experiments}
	% !TEX root = main.tex

\subsection{Benchmark suites and experimental setup}
We evaluate our solver on a representative selection of SMT-LIB benchmarks, including all instances from the SMT-LIB 2025 benchmark release\footnote{\url{https://zenodo.org/communities/smt-lib/}}. From the combined pool of roughly 100\,000 \verb|QF_S| and \verb|QF_SLIA| problems, we randomly sample 2\,000 instances proportional to each track’s share of the total, so that our test set reflects the underlying distribution of problem types.
The experiments were conducted on a MacBook Pro with 16 GB of RAM, running macOS Sonoma 14.5. The system was powered by an Apple M3 chip. The timeout for each experiment was set to 60 seconds.

\paragraph*{ECMAScript Regex Benchmarks.}
Our experimental evaluation in this paper does not include benchmarks making extensive use of ECMAScript-specific regular expression features such as look-arounds or capture groups.
A large-scale evaluation of this functionality was presented in~\cite{blackostrich}, using a benchmark set of approximately 8{,}800 ECMAScript-style regex patterns extracted from real-world web forms. 
The correctness of the semantics was validated in~\cite{popl22} by comparing OSTRICH’s results against JavaScript’s reference implementation. 
The ECMAScript regex component in OSTRICH2 is identical to that used in those earlier evaluations. 
Therefore, we omit repeating those experiments here and focus on broader SMT-LIB benchmarks.

\subsection{Performance evaluation}

The experiments are conducted on the following solvers: cvc5 1.2.0~\cite{cvc5}, Z3 4.15~\cite{z3}, Z3-alpha~\cite{z3alpha}, Z3-Noodler 1.3.0~\cite{noodler}, OSTRICH1 (the original OSTRICH)~\cite{ostrich}, and different engines of OSTRICH2\footnote{\url{https://doi.org/10.5281/zenodo.15378521}} and OSTRICH2 running the engines in a time-sliced portfolio using the flag.
For the RCP engine we evaluate two complementary configurations: one mode that combines forward and backward propagation without Nielsen splitting (using the flags \texttt{+F+B-N}), and another mode (using the flags \texttt{-F+B+N}) that disables forward propagation in order to emphasize Nielsen's equation decomposition. 
In practice, combining forward propagation and splitting yields limited synergy: the extra word equations generated by Nielsen splitting create significant propagation overhead, so omitting forward propagation often improves overall solving performance.  

\begin{figure}[htbp]
  
\centering
\resizebox{\linewidth}{!}{%
    \begin{tabular}{lccccc}
        \toprule
                        & cvc5 & Z3   & Z3-alpha & Z3-Noodler    & OSTRICH1 \\
        \midrule
        \textsc{sat}    & 1155 & 1097 & 1123     & \textbf{1201} & 825      \\
        \textsc{unsat}  & 729  & 728  & 728      & \textbf{744}           & 664      \\
        \midrule
        unknown/timeout & 116  & 175  & 149      & \textbf{55}            & 511      \\
        \midrule
        Solved          & 1884 & 1825 & 1851     & \textbf{1945}          & 1489     \\
        Total time [s]  &  7770 &  12158 &  10523     & \textbf{3660}  &  35711     \\
        \bottomrule
    \end{tabular}%
}
  \caption{The experiment result of cvc5, Z3, Z3-alpha, Z3-Noodler and OSTRICH1 on the SMT-LIB'25 benchmarks. Total time (in seconds) includes all instances, with each unknown/timeout counted as 60s.}
  \label{fig:experiment_res_z3_cvc5}
\end{figure}

\begin{figure}[htbp]
  
\centering
\resizebox{\linewidth}{!}{%
  \begin{tabular}{lccccc}
    \toprule
    \textbf{}       & OSTRICH2      & ADT  & CE   & RCP-F+B+N & RCP+F+B-N \\
    \midrule
    \textsc{sat}    & \textbf{1147}          & 610  & 970  & 1067      & 1042      \\
    \textsc{unsat}  & \textbf{758}  & 619  & 679  & 747       & 750       \\
    \midrule
    unknown/timeout & \textbf{95}   & 771  & 351  & 186       & 208       \\
    \midrule
    Solved          & \textbf{1905} & 1229 & 1649 & 1814      & 1792      \\
    Total time [s]  &  \textbf{14824} &  53689 &  24600     & 17680  &  19900     \\
    \bottomrule
  \end{tabular}%
}

  \caption{The experiment result of different engines of OSTRICH2 on the SMT-LIB'25 benchmarks.The flags \texttt{+F+B-N} enable forwards/backwards propagation and disable the NielsenSplitter. The flags \texttt{-F+B+N} disable forwards propagation and enable backwards propagation and the NielsenSplitter. Total time (in seconds) includes all instances, with each unknown/timeout counted as 60s.}
  \label{fig:experiment_res_ostrich}
\end{figure}

\noindent Figure~\ref{fig:experiment_res_z3_cvc5} shows that among the off‐the‐shelf solvers, Z3‐Noodler leads with 1\,945 solved problems, followed by cvc5 (1\,884), Z3-alpha (1\,851), Z3 (1\,825), and OSTRICH1 (1\,489).  Figure~\ref{fig:experiment_res_ostrich} shows that both RCP configurations (1\,814 and 1\,792 solves) and CE‐Str (1\,649) represent clear upgrades over OSTRICH1, whereas the ADT‐Str engine (1\,229) trails—largely due to its simple algorithm.
These per-engine gains can be attributed to improved preprocessing, an extended set of proof rules, and, for CE-Str, a different underlying solving technique.

In the full OSTRICH2 configuration, these upgraded engines are run in a time-sliced portfolio, which not only inherits the improvements of the individual modes but also exploits their complementary strengths, as each engine solves benchmarks that the others miss. This complementarity lifts OSTRICH2’s overall performance close to the top solvers in the field.

\noindent When we combine all four engines in a time‐sliced portfolio (15 seconds for each mode for a 60-second limit), OSTRICH2 solves 1\,905 problems, putting it just behind Z3‐Noodler and ahead of the rest of the field. 
The portfolio not only inherits the individual gains of each engine but also benefits from their complementary strengths, as each solves benchmarks the others miss. 
We further observe that Z3‐Noodler excels on the \textsc{sat} instances, while OSTRICH2 performs particularly well on the \textsc{unsat} cases.

Finally, although Z3‑Noodler does not support \texttt{replace\_all}, only 2.4\% of SMT‑LIB instances use this operator, so its absence has only a modest effect on overall rankings. 
Figure~\ref{fig:experiment_res_rep_all} shows the results on the benchmark set restricted to instances without \texttt{replace\_all}.
Among the 35 \texttt{replace\_all} benchmarks in our SMT-LIB sample, OSTRICH2 solves 25, while cvc5, Z3, and Z3-alpha each solve only two instances, and Z3-Noodler solves none.
This shows that most solvers provide limited support for this operator. 
When we exclude the \texttt{replace\_all} benchmarks, the rankings remain broadly unchanged; OSTRICH2 drops slightly below cvc5 in relative performance.

We remark that, since our evaluation uses a stratified random sample of 2\,000 instances drawn from the full pool of roughly 100\,000 \verb|QF_S| and \verb|QF_SLIA| benchmarks, there is inevitably some statistical variance in the exact per‐solver ranking; nonetheless, the fact that OSTRICH2 solves almost as many problems as the leader demonstrates its overall competitiveness.

\begin{figure}[htbp]
	
\centering
\resizebox{\linewidth}{!}{%
	\begin{tabular}{lccccc}
		\toprule
		& cvc5 & Z3   & Z3-alpha & Z3-Noodler    & OSTRICH2 \\
		\midrule
		\textsc{sat}    & 1155 & 1097 & 1123     & \textbf{1201} & 1141      \\
		\textsc{unsat}  & 727  & 726  & 726      & \textbf{744}           &  739      \\
		\midrule
		unknown/timeout & 83  & 142  & 116      & \textbf{20}           & 85     \\
		\midrule
		Solved          & 1882 & 1823 & 1849    & \textbf{1945}         &   1880   \\
        Total time [s]  &  5790 &  10174 &  8536     & \textbf{1617}  &  14070     \\		
		\bottomrule
	\end{tabular}%
}
  \caption{The experiment result of cvc5, Z3, Z3-alpha, Z3-Noodler and OSTRICH2 on the SMT-LIB'25 benchmarks \textbf{excluding} \texttt{replace\_all}. Total time (in seconds) includes all instances, with each unknown/timeout counted as 60s.}
	\label{fig:experiment_res_rep_all}
\end{figure}

	\section{Conclusion}\label{sec:conclusion}
	% !TeX root = main.tex

This paper has strived to provide a concise introduction of the string solver 
OSTRICH2, which inputs constraints in an extension of the SMT-LIB theory of 
Unicode Strings with transducers (through mutual recursion) and ECMAScript
regular expressions, among others. We have also empirically demonstrated its 
competitiveness with other string solvers on SMT-COMP benchmarks, particularly 
on unsatisfiable instances. Interested readers, who are interested in
participating in the development and/or applications of OSTRICH2, are
wholeheartedly encouraged to get in touch with us.

	\section*{Acknowledgements}

    We thank EPSRC [EP/T00021X/1], European Research Council (LASD, 101089343),
    and the  Swedish Research Council (grant 2021-06327) for their support.

	% References

	\bibliographystyle{IEEEtran}

	\bibliography{references}
	
	\appendix
	
	\section{SMT-LIB Grammar}
	
\section{Automata Definition}
\label{app:automata-grammar}

\[
\begin{array}{lcl}
	\textit{Automaton} &::=& \texttt{automaton}\; \textit{Ident}\; \texttt{\{}\; \texttt{init}\; \textit{State}\; \texttt{;}\\ &   &\textit{Tr}^*\; \texttt{accepting}\; \textit{State}\;(\texttt{,}\; \textit{State})^*\; \texttt{;}\; \texttt{\}}\;\\
	\textit{Tr} &::=& \textit{State}\; \texttt{->}\; \textit{State}\; \texttt{[}\; \textit{Int}\; \texttt{,}\; \textit{Int}\; \texttt{]}\; \texttt{;} \\
	\textit{State} &::=& \textit{Ident} \qquad\qquad \\
	\textit{Ident} &::=& [\texttt{A-Za-z0-9\_}]+ \\
	\textit{Int} &::=& [\texttt{0-9}]+
\end{array}
\]
\noindent Ranges are inclusive and denote Unicode code points as accepted by the current \texttt{re.from\_automaton} parser (0--65535)\footnote{While OSTRICH2 supports reasoning over the full Unicode range, the automaton parser is presently limited to BMP code points.}. States referenced in transitions are implicitly declared. Whitespace is insignificant.

\section{Transducer Definitions}
\label{app:transducer-grammar}

OSTRICH encodes \emph{transducers} as a collection of mutually recursive Boolean functions in SMT-LIB, where each function corresponds to a distinct state in the transducer.
A transducer consists of a set of such states that, at each step, inspect the first character(s) of the input and output strings, select a transition based on guarded conditions, and then recurse on the remaining substrings until termination.

Formally, the behaviour of each state function is structured as follows:
\begin{enumerate}
	\item Inspect the leading characters of the input and output strings using \texttt{str.head}.
	\item Evaluate guard conditions to determine the appropriate transition.
	\item Invoke another state function (possibly the same one) on the residual substrings obtained via \texttt{str.tail}.
\end{enumerate}

Acceptance is defined via \emph{base cases}, typically when both input and output have been fully consumed, expressed as \texttt{(= x "")} and \texttt{(= y "")}.
Character comparisons are performed using Unicode code points, with numeric ranges used to express character classes (e.g., lowercase letters correspond to codes $97$--$122$).
\subsection{Example 1: General Template}
In the template below, the transducer is defined by two mutually recursive state functions, \texttt{S} and \texttt{S2}, each taking as arguments the remaining portions of the input (\texttt{x}) and output (\texttt{y}) strings. The first clause of each state specifies the \emph{base case}, here accepting when both \texttt{x} and \texttt{y} are empty. 
Subsequent clauses correspond to guarded transitions: in \texttt{S}, the first transition matches and deletes an \texttt{'a'} from the input (consuming only \texttt{x} and leaving \texttt{y} unchanged), while the second transition copies the current character from input to output (consuming one symbol from each).
Each transition invokes the appropriate successor state on the residual strings obtained via \texttt{str.tail}. The second state, \texttt{S2}, is structurally similar, here illustrated with only a copying transition. This structure generalises to arbitrary transducers by adding states, refining guard conditions, and controlling which side(s) of the input–output pair are consumed in each branch.

\begin{lstlisting}[language=SMTLib,frame=single]
(set-option :parse-transducers true)
; One state per function, (x,y) are the remaining input/output.
(define-funs-rec
  ((S ((x String) (y String)) Bool)
    (S2 ((x String) (y String)) Bool))
  (
   ; S: base case = accept when both consumed
    (or (and (= x "") (= y ""))
   ; transition 1: delete 'a' from input
       (and (not (= x ""))
         (= (str.head x) (char.from-int 97)) ; 'a'
         (S (str.tail x) y))
       ; transition 2: copy char (consume both)
       (and (not (= x "")) (not (= y ""))
         (= (str.head x) (str.head y))
         (S (str.tail x) (str.tail y))))
    ; S2: another state if needed...
    (or (and (= x "") (= y ""))
        (and (not (= x "")) (not (= y ""))
        (= (str.head x) (str.head y))
        (S2 (str.tail x) (str.tail y))))
  )
\end{lstlisting}
\subsection{Example 2: \texttt{toUpper}}
The \texttt{toUpper} transducer illustrates a length-preserving character transformation. It consists of a single state, \texttt{toUpper}, which accepts when both input \texttt{x} and output \texttt{y} are empty. Otherwise, the guard requires that both strings be non-empty, and the head character of the output is constrained to be either the uppercase equivalent of the head of the input (if the input character is a lowercase letter, identified by Unicode codes $97$–$122$) or an exact copy of the input character in all other cases. This transformation is expressed using the \texttt{ite} construct, subtracting $32$ from the character code to obtain the uppercase form when applicable. The function then recurses on the tails of both strings, ensuring that the transformation is applied position-wise until the entire input has been processed.

\begin{lstlisting}[language=SMTLib,frame=single]
(set-option :parse-transducers true)

(define-fun-rec toUpper ((x String) (y String)) Bool
(or (and (= x "") (= y ""))
  (and (not (= x "")) (not (= y ""))
     (= (char.code (str.head y))
       (ite (and (<= 97 (char.code (str.head x)))
                 (<= (char.code (str.head x)) 122))
            (- (char.code (str.head x)) 32)
            (char.code (str.head x))))
     (toUpper (str.tail x) (str.tail y)))))
\end{lstlisting}
\subsection{Example 3: \texttt{extract1st}}
The \texttt{extract1st} transducer demonstrates a non-length-preserving transformation involving multiple states. 
Its purpose is to scan the input \texttt{x} for the first occurrence of the character \texttt{'='} (Unicode code $61$), copy the subsequent characters into the output \texttt{y} until the next \texttt{'='} is encountered, and then skip the remainder of the input. 
The initial state, \texttt{extract1st}, advances through the input without producing output until it finds the first \texttt{'='}, at which point it transitions to \texttt{extract1st\_2}. 
In this second state, the transducer consumes characters from both input and output in lockstep, copying them directly unless another \texttt{'='} is reached, which triggers a transition to \texttt{extract1st\_3}. 
The final state, \texttt{extract1st\_3}, consumes the remaining input without producing further output, thereby terminating the extraction. 
For example, given the input string \texttt{x = "foo=bar"}, the transducer outputs \texttt{y = "bar"}.
\begin{lstlisting}[language=SMTLib,frame=single]
(set-option :parse-transducers true)
	
(define-funs-rec ((extract1st ((x String) (y String)) Bool)
  (extract1st_2 ((x String) (y String)) Bool)
  (extract1st_3 ((x String) (y String)) Bool)) (
  ;
  ; extract1st
  (or (and (= x "") (= y ""))
    (and (not (= x ""))
      (not (= (str.head x) (char.from-int 61)))  ; not '='
      (extract1st (str.tail x) y))
    (and (not (= x ""))
      (= (str.head x) (char.from-int 61))        ; '='
      (extract1st_2 (str.tail x) y)))
  ;
  ; extract1st_2
  (or (and (= x "") (= y ""))
    (and (not (= x "")) (not (= y ""))
      (= (str.head x) (str.head y))
      (not (= (str.head x) (char.from-int 61)))  ; not '='
      (extract1st_2 (str.tail x) (str.tail y)))
    (and (not (= x ""))
      (= (str.head x) (char.from-int 61))        ; '='
      (extract1st_3 (str.tail x) y)))
  ;
  ; extract1st_3
  (or (and (= x "") (= y ""))
    (and (not (= x ""))
      (extract1st_3 (str.tail x) y)))
))
\end{lstlisting}

\subsection{Grammar for OSTRICH transducers}
The following grammar summarises the fragment of SMT-LIB used by OSTRICH for defining such transducers.

{\footnotesize\ttfamily
\[
\begin{array}{rcl}
	Transducer &::=& 
	\texttt{(define-fun-rec} \; StateSig \; Body \texttt{)}
	\\
	&\mid&
	\texttt{(define-funs-rec} \; (StateSig^{+}) \; (Body^{+}) \texttt{)}
	\\[6pt]
	
	StateSig &::=& 
	\texttt{(} Id \; \texttt{((x String) (y String)) Bool} \texttt{)}
	\\[6pt]
	
	Body &::=& 
	\texttt{(or} \; Clause^{+} \texttt{)}
	\\[6pt]
	
	Clause &::=& 
	BaseCase
	\;\mid\;
	Transition
	\\[6pt]
	
	BaseCase &::=& 
	\texttt{(and} \; Guard^{*} \texttt{)} 
	\\[6pt]
	
	Transition &::=& 
	\texttt{(and} \; Guard^{+} \; Call \texttt{)}
	\\[6pt]
	
	Call &::=& 
	\texttt{(} Id \; XArg \; YArg \texttt{)}
	\\[6pt]
	
	XArg &::=& 
	\texttt{x} 
	\;\mid\; 
	\texttt{(str.tail x)}
	\\[6pt]
	
	YArg &::=& 
	\texttt{y} 
	\;\mid\; 
	\texttt{(str.tail y)}
	\\[6pt]
	
	Guard &::=& 
	\texttt{(=} \; Term \; Term \texttt{)}
	\\
	&\mid& \texttt{(not} \; Guard \texttt{)}
	\\
	&\mid& \texttt{(and} \; Guard^{+} \texttt{)}
	\\
	&\mid& \texttt{(or} \; Guard^{+} \texttt{)}
	\\
	&\mid& \texttt{(<=} \; Int \; Int \texttt{)}
	\\[6pt]
	
	Term &::=& 
	Char
	\;\mid\;
	\texttt{(str.head x)}
	\;\mid\;
	\texttt{(str.head y)}
	\\
	&\mid& \texttt{(char.code} \; Term \texttt{)}
	\\
	&\mid& \texttt{(ite} \; Guard \; Term \; Term \texttt{)}
	\\[6pt]
	
	Char &::=& 
	\texttt{(char.from-int} \; Int \texttt{)}
	\\
	&\mid& \texttt{(str.head x)}
	\\
	&\mid& \texttt{(str.head y)}
	\\[6pt]
	
	Int &::=& 
	Numeral
	\;\mid\;
	\texttt{(char.code} \; Term \texttt{)}
	\\
	&\mid& \texttt{(+} \; Int \; Int \texttt{)}
	\\
	&\mid& \texttt{(-} \; Int \; Int \texttt{)}
	\\
\end{array}
\]
}
In this grammar, each \texttt{StateSig} corresponds to one \emph{state} in the transducer. 
The \texttt{Body} of a state consists of one or more \texttt{Clauses} combined with \texttt{or}, where each clause represents either:
\begin{itemize}
	\item a \emph{base case}, such as \texttt{(and (= x "") (= y ""))}, or
	\item a \emph{transition} guarded by conditions on the heads of \texttt{x} and \texttt{y} and leading to a recursive \texttt{Call} on their tails.
\end{itemize}
The \texttt{Guard} syntax captures conditions such as character equality, inequality, or membership in a Unicode range (as in \texttt{toUpper}, which checks lowercase via \texttt{(<= 97 (char.code (str.head x)))} and \texttt{(<= (char.code (str.head x)) 122)}). 
\texttt{XArg} and \texttt{YArg} indicate whether a transition consumes from the input, output, or both. This is the key distinction between length-preserving transformations (both consumed) and non-length-preserving ones (only one consumed at a time).

\end{document}